\documentclass[pra,notitlepage,twocolumn,amsmath,amssymb,aps,superscriptaddress]{revtex4-2}

\usepackage{amsmath,amssymb,amsfonts,bbm,times}
\usepackage[dvipsnames]{xcolor}
\usepackage[T1]{fontenc}
\usepackage{braket}
\usepackage[normalem]{ulem}
\usepackage{graphicx}
\usepackage{bm}
\usepackage{physics}

 \usepackage{soul}

\usepackage[colorlinks=true,bookmarks=false,linkcolor=RoyalBlue,urlcolor=RoyalBlue,citecolor=RoyalBlue,breaklinks]{hyperref}

\begin{document}

\title{Quantum observables over time for information recovery}

\author{Gabriele Bressanini}
\affiliation{Blackett Laboratory, Imperial College London, London SW7 2AZ, United Kingdom}
\author{Farhan Hanif}
\affiliation{Blackett Laboratory, Imperial College London, London SW7 2AZ, United Kingdom}
\author{Hyukjoon Kwon}
\affiliation{Korea Institute for Advanced Study, Seoul 02455, South Korea}
\author{M.S. Kim}
\affiliation{Blackett Laboratory, Imperial College London, London SW7 2AZ, United Kingdom}
\affiliation{Korea Institute for Advanced Study, Seoul 02455, South Korea}

\begin{abstract}
We introduce the concept of quantum observables over time (QOOT), an operator that jointly describes two observables at two distinct time points, as a dual of the quantum state over time formalism. 
We provide a full characterization of the conditions under which a QOOT can be properly defined, via a no-go theorem. 
We use QOOTs to establish a notion of time-reversal for generic quantum channels with respect to a reference observable, enabling the systematic construction of recovery maps that preserve the latter.
These recovery maps, although generally non-physical, can be decomposed into realizable channels, enabling their application in noiseless expectation value estimation tasks.
We provide explicit examples and compare our protocol with other error mitigation methods.
% We show that our approach can achieve optimal performance, measured in terms of sampling overhead, outperforming probabilistic error cancellation.
We show that our protocol retrieves the noiseless expectation value of the reference observable and can achieve optimal sampling overhead, outperforming probabilistic error cancellation.
\end{abstract}

\maketitle

\section{Introduction}
Due to unavoidable interactions with the environment, realistic quantum systems are inherently noisy, resulting in their evolution being governed by non-unitary quantum channels \cite{nielsen2010quantum, wilde2013quantum}.
Consequently, any information we extract from the system is also corrupted by noise.
A fundamental question that arises is then how to retrodict the initial state of a quantum system that has undergone such noisy dynamics.
In particular, one might ask whether there exists a physical process (i.e., a quantum channel) that, starting from a state affected by noise, can recover the original state by undoing the effects of noise. 
However, in general, this is feasible only for unitary channels, as they are the only quantum channels that are reversible.
Strictly speaking, the notion of time reversal in quantum theory is well-defined for unitary evolutions only, that is for closed systems' dynamics.
%A fundamental question that arises is how to undo the effects of noise on a quantum state that has undergone such noisy dynamics.
%One might ask whether it is possible to reverse this noise through an inverse quantum process.
%However, in general, this is only achievable for unitary channels, as they are the only quantum channels that are invertible.
%Strictly speaking, the notion of time reversal in quantum theory is well-defined for unitary evolutions only, i.e. for close systems' dynamics.
A way to overcome this limitation is to introduce an additional element: a reference state, or prior, with respect to which the inversion is performed, along with the noise channel \cite{fullwood2022royal,parzygnat2023axioms}. This leads to the introduction of a robust notion of time-reversal for generic quantum channel describing open quantum system \cite{QuantumBayesRetrodiction}, in analogy with classical retrodiction.

In this framework, the reverse process is defined as a quantum map that only guarantees the recovery of the reference state.
For instance, the well known Petz recovery map \cite{petz1986sufficient,petz_continuous,barnum2002reversing,wilde2015recoverability} is a quantum channel that satisfies the aforementioned recovery property.
If one further relaxes the requirement that the reverse process must be a quantum channel, other possible definitions of an inverse process for a given noise channel and reference state become possible. Of course, if we are willing to relax the complete-positivity condition, the mathematical inverse of a channel can serve this purpose \cite{temme_error_mitigation}. However, this inverse may not always be defined, nor be the most suitable choice, as we will discuss later.

The notion of a quantum state over time (QSOT) was introduced as the quantum analogue of a joint probability distribution for a bipartite system across two time-like separated regions \cite{leifer2013towards,horsman2017can}.
There are multiple ways to define states over time \cite{fullwood2022royal,horsman2017can,leifer2013towards,ohya1983note,ohya1983compound,asorey2006relations, Lie2024quantum} because, while the postulates of quantum mechanics specify how to describe joint states in space-like separated regions, they provide no guidance on how to jointly represent states that are time-like separated.
In general, an appropriate definition of QSOT should be an operator acting on the tensor product of the Hilbert space at two different times and, if the two time-like separated states are causally related, then the QSOT should encode such correlations as well.
Furthermore, as highlighted in Refs. \cite{fullwood2022royal,fitzsimons2015quantum}, for a QSOT to capture temporal correlations it must allow for negative eigenvalues, meaning that states over time are not positive operators in general.
The authors of Ref.~\cite{QuantumBayesRetrodiction} then used the QSOT formalism and a notion of time-reversal symmetry in classical systems to systematically define Bayes rule in the quantum framework.
This approach results in a new definition of ``Bayesian inverse'' of a quantum channel with respect to a reference state. Crucially, these maps inherently possess the recovery property of the reference state we have mentioned earlier.
Naturally, different choices of QSOT lead to different recovery maps. For example, the Leifer-Spekkens QSOT \cite{LeiferSpekkens1,LeiferSpekkens2} leads to a Bayesian inverse that corresponds to the Petz recovery map. 

So far, most research efforts have been focused on defining retrodiction maps for a given noise channel and reference state. 
In this paper, we shift perspective and turn our attention to quantum observables. Our goal is to find recovery maps that protect specific observables from the effects of a given noisy dynamics. 
From a more operational standpoint, such maps would allow us to devise protocols that preserve the expectation value of a specific observable against a given noise model, \emph{for every possible state}.
This is in contrast with state retrodiction, which cannot guarantee to perfectly infer the expectation values of observables beyond the reference state.
This shift in perspective is motivated by the fact that several quantum information protocols and quantum algorithms rely on extracting information through expectation values, which are inevitably impacted by the noise affecting the system. Quantum error mitigation addresses the problem of protecting expectation values from noisy dynamics.
In this context, unphysical recovery maps do not pose a problem, as they can be decomposed into physically-realizable quantum channels \cite{buscemi2013direct,physical_implementability_linear_maps,temme_error_mitigation}, which further justifies the extension of the search for recovery maps to non-CP maps.
This quasi-probability decomposition then serves as a simulation tool to estimate the noise-free expectation value, following the same strategy used in the well-known probabilistic error cancellation (PEC) \cite{temme_error_mitigation,practical_error_mitigation} technique, as we will explore in more detail later.
In Ref.~\cite{information_recoverability}, the authors address the problem of constructing recovery maps for specific observables under a given noise model using a different approach. The main difference between our methods lies in the fact that, while their approach requires solving a challenging optimization problem, we take a more fundamental route, leveraging time-reversal symmetries to systematically derive the recovery maps.

This paper has two primary objectives. First, from a foundational perspective, we aim to generalize the state over time formalism to encompass quantum observables, by introducing the concept of a quantum observable over time (QOOT). 
The latter is an operator that jointly describes two quantum observables that are time-like separated, while also capturing their temporal correlations. Unlike the QSOT, we show that the construction of the QOOT is not universally applicable. 
Specifically, we fully characterize the pairs of quantum channels and reference observables for which the QOOT can be properly defined.
Second, we leverage observables over time to implicitly and systematically define recovery maps via an equation that reflects a notion of time-reversal symmetry, thus offering a definition of time-reversal for quantum channels with respect to a reference observable.
On a more applied front, inspired by the PEC technique, we investigate the potential of these recovery maps for quantum error mitigation, focusing on the task of estimating the noise-free expectation value of a specific reference observable subjected to a known noise channel.

The rest of this paper is organized as follows.
In Sec.~\ref{sec_QOOT}, we introduce the concept of quantum observable over time through its defining properties and present a necessary and sufficient condition under which such an object can be properly defined.
In Sec.~\ref{sec_recovery_maps}, we demonstrate how observables over time may be used to write an equation that expresses time-reversal symmetry of the system, leading to the implicit definition of recovery maps for both pre- and post-processing protocols.
In Sec.~\ref{sec_jordan_product}, we introduce a specific instance of the quantum observable over time, based on the Jordan product, that will be used throughout the paper due to its desirable properties. For this choice of QOOT, we then solve the equations, obtain the corresponding recovery maps and study their properties. In Sec.~\ref{sec_simulating_recovery_maps}, we show how an unphysical recovery map can be expressed in terms of physically-implementable maps and discuss the application of this decomposition in noiseless expectation value estimation tasks. We then highlight differences and similarities between this approach and established error mitigation protocols. In Sec.~\ref{sec_examples}, we explicitly compute recovery maps for relevant noise models, we decompose them in terms of physically-realizable maps, and discuss their implementation cost, comparing it with known optimality results. Lastly, in Sec.~\ref{sec_conclusions} we draw conclusions, highlight remaining open questions and offer final remarks.

\section{Quantum observables over time}
\label{sec_QOOT}
In this Section, we introduce the quantum observable over time, i.e., the fundamental object that will be used throughout the paper to obtain recovery maps. Our construction follows that of quantum states over time presented in Ref.~\cite{QuantumBayesRetrodiction}. 
Let us consider an observable (i.e., a Hermitian operator) $O$ and a Hermitian-preserving (HP) map $\mathcal{E}$ acting on the space of states as $\mathcal{E}: \rho \mapsto \mathcal{E}(\rho)$ (Schr\"{o}dinger-like evolution).
Equivalently, the adjoint map $\mathcal{E}^\dag$ describes the dynamical evolution of observables (Heisenberg-like evolution), namely $\mathcal{E}^\dag : O \mapsto \mathcal{E}^\dag (O)$. Here, $O$ acts on the Hilbert space $\mathcal{H}_A$ and $\mathcal{E}^\dag (O)$ acts on $\mathcal{H}_B$.
Equivalently, we write $O\in\mathcal{L}(\mathcal{H}_\mathcal{A})$ and $\mathcal{E}^\dag (O)\in\mathcal{L}(\mathcal{H}_\mathcal{B})$, where $\mathcal{L}(\mathcal{H})$ denotes the space of linear operators acting on the Hilbert space $\mathcal{H}$.
%\hk{(HK: Shouldn't it be reversed? note that $\tr[O{\cal E}(\rho)] = \tr[{\cal E}^\dagger(O) \rho] $)}
The Schr\"{o}dinger and Heisenberg pictures are well-known to provide completely equivalent descriptions of a quantum system undergoing dynamical evolution. This duality becomes evident when considering the expectation value of an observable $O$ for a system initially in state $\rho$, evolving under the map $\mathcal{E}$. Specifically, the expectation value can be expressed equivalently as
\begin{equation}
    \Tr[\mathcal{E} (\rho) O] = \Tr[ \rho\, \mathcal{E}^\dagger (O)] \, .
\end{equation}
We seek to define a \emph{quantum observable over time},  $\mathcal{E}^\dag \star O$, i.e. an operator acting on $\mathcal{H}_A \otimes \mathcal{H}_B$ that can jointly describe the reference observable $O$ and its dynamically evolved counterpart $\mathcal{E}^\dag (O)$. In particular, such an object must correctly retrieve the marginal observables, namely it should satisfy the following minimal requirements
\begin{equation}
    \Tr_{B}[\mathcal{E}^\dagger \star O] = O \, ,
    \label{minimal_property_1}
\end{equation}
\begin{equation}
    \Tr_{A}[\mathcal{E}^\dagger \star O] = \mathcal{E}^\dagger (O) \, .
    \label{minimal_property_2}
\end{equation}
Additional, desirable properties for $\mathcal{E}^\dag \star O$ may be linearity in the reference observable, process-linearity, and Hermitianity.
We point out that the Leifer-Spekkens QSOT, which gives rise to the Petz recovery map, can not be promoted to a QOOT. This is because its definition involves taking the square root of the reference state. However, when the reference is a generic quantum observable, we cannot ensure its positivity.

Interestingly, we show via a no-go theorem that, for a generic HP map $\mathcal{E}$ and quantum observable $O$, it is not always possible to construct a QOOT, i.e., an operator acting on $\mathcal{H}_A \otimes \mathcal{H}_B$ satisfying the minimal partial trace conditions outlined above. This stands in sharp contrast to the QSOT formalism, where such a construction is always possible for any pair of reference state and quantum channel.
We can easily see this by taking the partial trace over $\mathcal{H}_B$ ($\mathcal{H}_A$) of Eq.~\eqref{minimal_property_1} (Eq.~\eqref{minimal_property_2}), which leads to 
\begin{equation}
    \label{no_go_condition}
    \Tr [\mathcal{E}^\dag \star O] = \Tr [O] = \Tr [\mathcal{E}^\dag (O)] \, .
\end{equation}
Hence, a necessary and sufficient condition for the QOOT to be well-defined is that the trace of $\mathcal{E}^\dag (O)$ coincides with that of $O$. Notice that this does not automatically imply that the map $\mathcal{E}^\dag$ needs to be unital, since the above condition needs to only be satisfied for the reference observable $O$.
In the following, we will assume that the combinations of maps and observables are such that QOOTs can be constructed. 
We note that, assuming Eq.~\eqref{no_go_condition} holds, the following trivial expression defines a valid QOOT: 
\begin{equation}
    \label{uncorrelate_OOT}
    \mathcal{E}^\dag \star O = \frac{O \otimes \mathcal{E}^\dag (O)}{\Tr[O]}  \, ,
\end{equation}
referred to as the \emph{uncorrelated} quantum observable over time. However, this form has limited usefulness, as we will demonstrate in the next Section, since it fails to encode any temporal correlations and treats $O$ and $\mathcal{E}^\dag (O)$ independently.

\section{Time-reversal symmetry and quantum observable retrodiction}
\label{sec_recovery_maps}
In this Section, we show how observables over time may be employed to implicitly define recovery maps via an equation that expresses time-reversal symmetry and resembles, in spirit, the Bayes rule and its quantum generalizations \cite{QuantumBayesRetrodiction}.
Let us consider a completely-positive trace-preserving (CPTP) map $\mathcal{E}:\mathcal{L}(\mathcal{H}_\mathcal{B})\rightarrow \mathcal{L}(\mathcal{H}_\mathcal{A})$ that models a noisy process, and a quantum observable $O\in\mathcal{L}(\mathcal{H}_\mathcal{A})$ we wish to protect.
We anticipate that, despite the limitations imposed by Eq.~\eqref{no_go_condition}, the formalism developed in this paper is able to generate recovery maps for physically-relevant scenarios, such as the protection of a Pauli observable under generalized amplitude damping or stochastic Pauli noise.

For a pre-processing protocol, we seek a HP map $\mathcal{P}:\mathcal{L}(\mathcal{H}_\mathcal{A})\rightarrow \mathcal{L}(\mathcal{H}_\mathcal{B})$ that satisfies the following (pre-processing) recovery property 
%\hk{(HK: One possible criticism would be that `pre-processing' might not be regarded as 'recovery operation,' as recovery is usually applied after noise. In a similar vein, it would be better to use different symbols for pre- (e.g., ${\cal P})$ and post-processing. Sorry for raising these points too late. However, if it takes too much effort---for example, we need to change Sec. V accordingly---I don't mind leaving it as it is.)}
\begin{equation}
    O = \mathcal{P}^\dag (\mathcal{E}^\dag (O)) \, .
    \label{pre_processing_recovery_property}
\end{equation}
\begin{figure}
    \centering
    \includegraphics[width=0.3\textwidth]{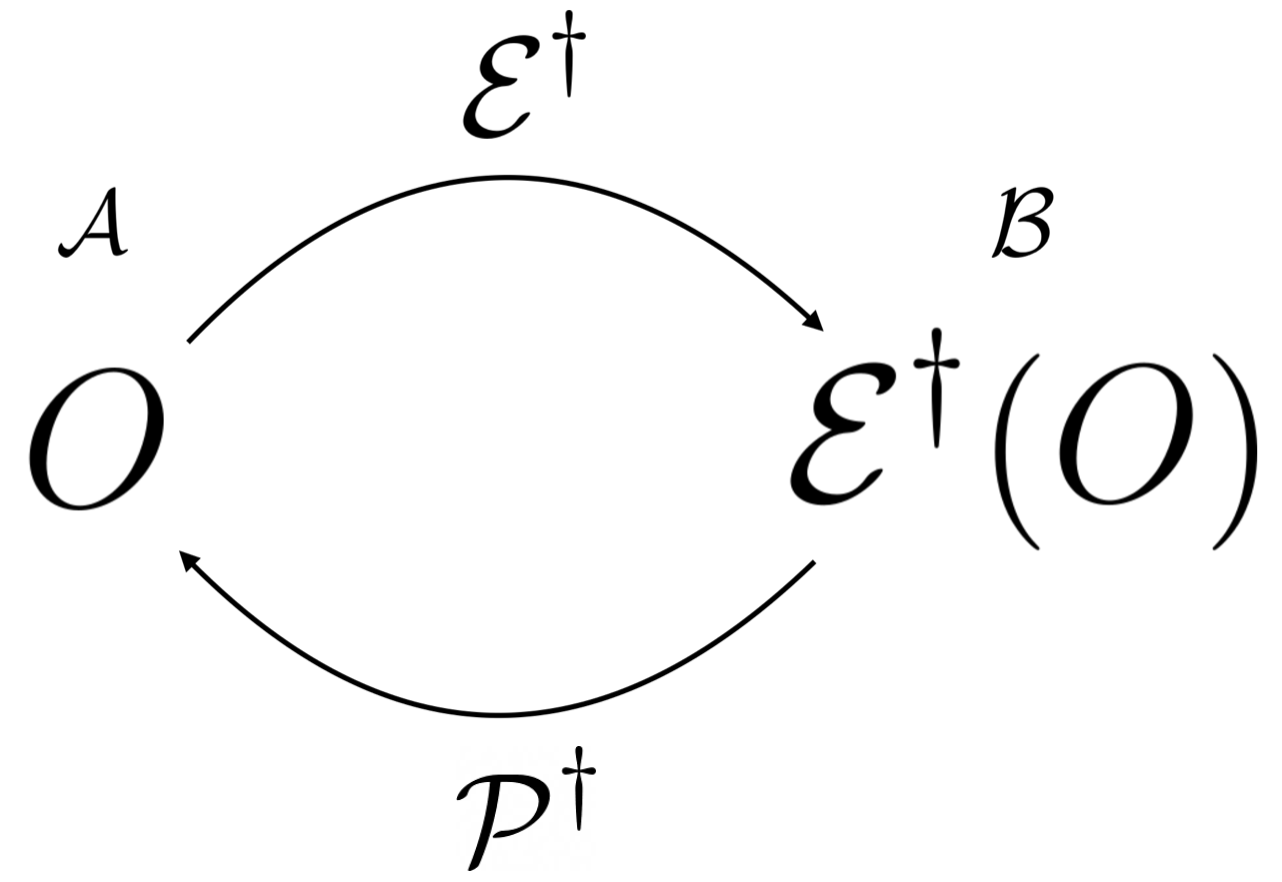}
    \caption{Pictorial representation of the pre-processing protocol.
    Depending on which observable we take as reference to build the QOOT (time-reversal symmetry), the system may be equivalently described by $\mathcal{E}^\dag \star O$ or by $\mathcal{P}^\dag \star \mathcal{E}^\dag (O)$.
    The pre-processing map acts on the noisy observable to retrieve the target observable $O$, i.e. $\mathcal{P}^\dag (\mathcal{E}^\dag (O)) = O$.}
    \label{fig_pre_processing}
\end{figure}
The expression above leads to a \emph{pre-processing} protocol because the action of $({\mathcal{P}^\dag} \circ {\mathcal{E}^\dag})$ on the observable $O$ corresponds to the action of $\left( \mathcal{E} \circ \mathcal{P} \right)$ on the state in Schr\"odinger picture. 
Hence, the pre-processing map $\mathcal{P}$ acts on the input state \emph{before} it undergoes the noisy evolution.
%\begin{equation}
%    \Tr[\rho O] = \Tr[\rho \mathcal{R}^\dag (\mathcal{E}^\dag (O))] = \Tr[\mathcal{E}(\mathcal{R}(\rho)) O ] \, ,
%\end{equation}
We can implicitly define such a map via the following equation, which reflects a certain time-reversal symmetry \cite{QuantumBayesRetrodiction} we wish to impose (see Fig.~\ref{fig_pre_processing} for a schematic representation)
\begin{equation}
    \mathcal{E}^\dag \star O = \tau (\mathcal{P}^\dag \star \mathcal{E}^\dag (O)) \, .
    \label{bayes_rule_pre_processing}
\end{equation}
Here $\tau: \mathcal{L}(\mathcal{H}_B) \otimes \mathcal{L}(\mathcal{H}_A) \rightarrow \mathcal{L}(\mathcal{H}_A) \otimes \mathcal{L}(\mathcal{H}_B)$ is the quantum-time-reversal map for observables over time, and it is defined as the unique conjugate-linear extension of 
\begin{equation}
    \tau (B\otimes A) = A^\dag \otimes B^\dag \, .
    \label{quantum_time_reversal_operator}
\end{equation}
The operator-swap introduced by $\tau$ places the two observables over time appearing in Eq.~\eqref{bayes_rule_pre_processing} on an equal footing: this is similar to what happens in the classical setting of the Bayes rule, where two joint probability distributions describing the same system are equated via time-reversal \cite{QuantumBayesRetrodiction}.
Additionally, the map $\tau$ applies Hermitian conjugation, a feature that was shown to make the notion of a time-reversal map in the quantum setting more robust \cite{QuantumBayesRetrodiction}.
By taking the partial trace over Hilbert space $\mathcal{H}_B$ on both sides of Eq.~\eqref{bayes_rule_pre_processing} we clearly obtain the pre-processing recovery property Eq.~\eqref{pre_processing_recovery_property}. This implies that a solution to Eq.~\eqref{bayes_rule_pre_processing} always satisfies the recovery property for the system at study, provided that the two observables over time appearing in the equation are well-defined, namely when  $\Tr[O] = \Tr[\mathcal{E}^\dag(O)] = \Tr[\mathcal{P}^\dag (\mathcal{E}^\dag (O))]$. We will come back to this point in detail later, after choosing a specific functional form for the QOOT.
The physical intuition behind Eq.~\eqref{bayes_rule_pre_processing} is that we may describe the same joint system using two observables over time: one for the ``forward'' process $\mathcal{E}^\dag$, using $O$ as the reference observable, and one for the ``backward'' process $\mathcal{P}^\dag$ that instead utilizes the noisy observable $\mathcal{E}^\dag (O)$ as reference operator.
Note that if we use the uncorrelated QOOT from Eq.~\eqref{uncorrelate_OOT}, Eq.~\eqref{bayes_rule_pre_processing} reduces to 
\begin{equation}
    O \otimes \mathcal{E}^\dag (O) = \mathcal{P}^\dag (\mathcal{E}^\dag (O)) \otimes \mathcal{E}^\dag (O) \, ,
\end{equation}
which is solved by \emph{any} map $\mathcal{P}$ such that $\mathcal{P}^\dag (\mathcal{E}^\dag (O)) = O$. However this formulation does not indicate \emph{which} of these recovery maps we should chose, hence it does not provide a robust definition of recovery map.

\begin{figure}
    \centering
    \includegraphics[width=0.3\textwidth]{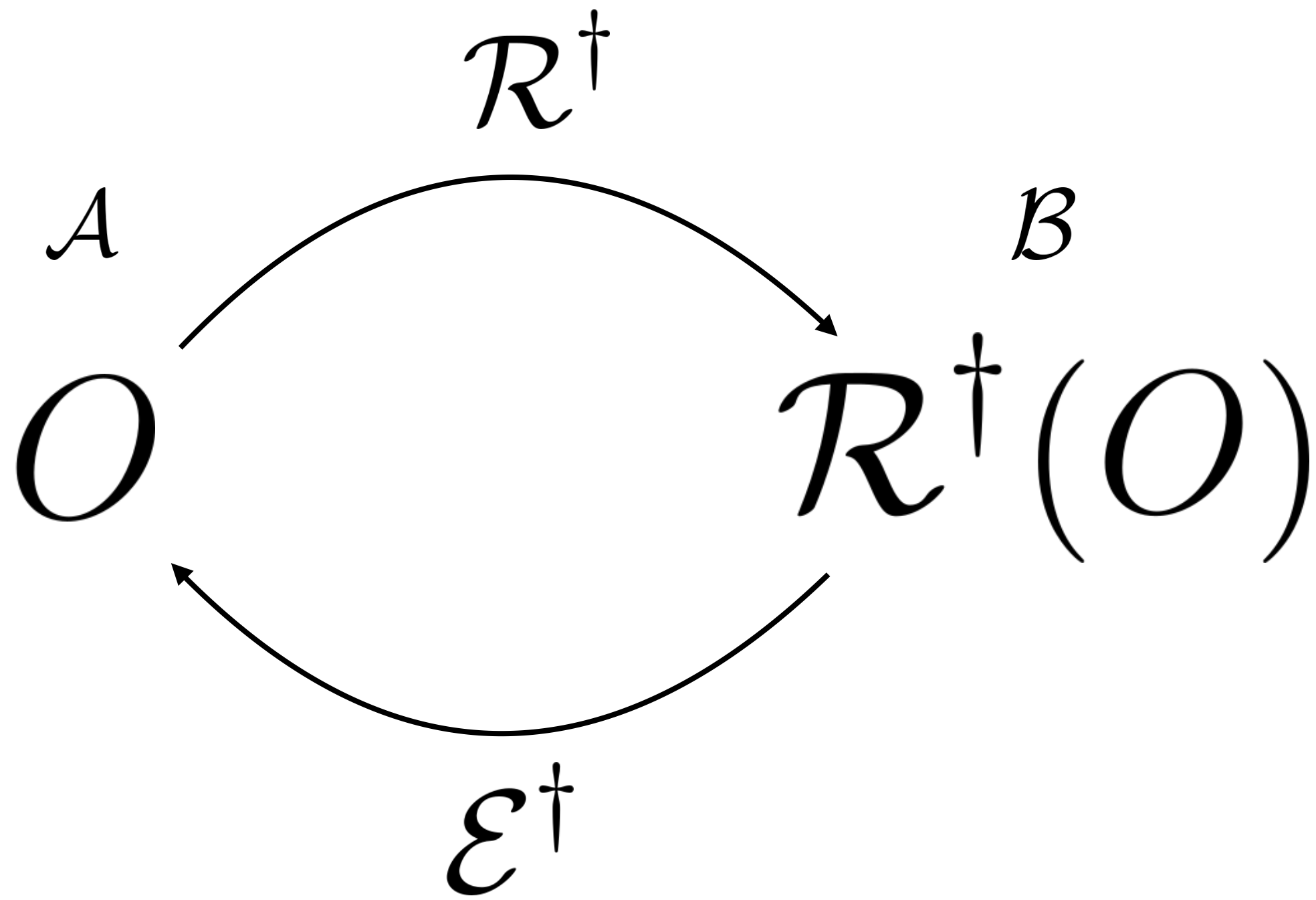}
    \caption{Pictorial representation of the post-processing protocol.
    Depending on which observable we take as reference to build the QOOT (time-reversal symmetry), the system may be equivalently described by $\mathcal{R}^\dag \star O$ or by $\mathcal{E}^\dag \star \mathcal{R}^\dag (O)$.
    The recovery map acts on the target observable $O$ so that the subsequent action of the noise map yields target observable itself, i.e. $\mathcal{E}^\dag (\mathcal{R}^\dag (O)) = O$.}
    \label{fig_post_processing}
\end{figure}

Analogously, for a post-processing protocol we seek a recovery map $\mathcal{R}:\mathcal{L}(\mathcal{H}_\mathcal{B})\rightarrow \mathcal{L}(\mathcal{H}_\mathcal{A})$ that satisfies the (post-processing) recovery property
\begin{equation}
    O = \mathcal{E}^\dag (\mathcal{R}^\dag (O)) \, .
    \label{recovery_property_post_processing}
\end{equation}
The expression above results in a \emph{post-processing} protocol, as the action of $({\mathcal{E}^\dag} \circ {\mathcal{R}^\dag})$ on the observable $O$ corresponds to the action of $\left( \mathcal{R} \circ \mathcal{E} \right)$ on the state in Schr\"odinger picture. 
Consequently, the recovery map $\mathcal{R}$ acts on the state \emph{after} it has undergone the noisy dynamics.
Such a recovery map may be implicitly defined via the following equation (see Fig.~\ref{fig_post_processing} for a schematic representation) 
\begin{equation}
    \mathcal{R}^\dag \star O = \tau (\mathcal{E}^\dag \star \mathcal{R}^\dag (O)) \, ,
    \label{bayes_rule_post_processing}
\end{equation}
which, once again, expresses a time-reversal symmetry we impose on the joint system. 
Notice how the introduction of $\mathcal{R}^\dag (O)$, i.e., the pre-processed target observable via the (adjoint) recovery map, is necessary in order to obtain a post-processing protocol.
Taking the partial trace of both sides of Eq.~\eqref{bayes_rule_post_processing} over $\mathcal{H}_B$ yields the recovery property Eq.~\eqref{recovery_property_post_processing}, provided that $\Tr[O] = \Tr[\mathcal{R}^\dag (O)] = \Tr [\mathcal{E}^\dag (\mathcal{R}^\dag (O))]$.

\section{Jordan product quantum observable over time}
\label{sec_jordan_product}
Previously, we introduced the quantum observable over time object in full generality and showed how it can be used to write equations Eq.~\eqref{bayes_rule_pre_processing} and Eq.~\eqref{bayes_rule_post_processing} that reflect a particular time-reversal symmetry of the system, and implicitly define recovery maps that protect the target observable $O$ from noise modeled by the CPTP map $\mathcal{E}$.
In this Section, we present a specific instance of the quantum observable over time that will be used throughout the paper to derive recovery maps. 
To this end, we first introduce the Jamio\l{}kowsi state \cite{jamiolkowski1972linear} of a quantum map $\mathcal{E}$, namely
\begin{equation}
    \label{jamiolkowsi_state}
    \mathcal{D}[\mathcal{E}] = \sum_{ij} \ketbra{i}{j} \otimes \mathcal{E}(\ketbra{j}{i}) \, ,
\end{equation}
where $\lbrace \ket{i} \rbrace$ denotes a generic basis of the Hilbert space. Given a HP map $\mathcal{E}^\dag : \mathcal{L}(\mathcal{H}_A) \rightarrow  \mathcal{L}(\mathcal{H}_B) $ and a reference observable $O$ acting on $\mathcal{H}_A$, we define the \emph{Jordan product quantum observable over time} as 
\begin{equation}
    \label{jordan_product_QOOT_definition}
    \mathcal{E}^\dag \star O = \frac{1}{2} \lbrace O \otimes \mathcal{I}, \mathcal{D}[\mathcal{E}^\dag] \rbrace \,  ,
\end{equation}
where $\mathcal{I}$ denotes the identity operator and the curly brackets denote the usual anti-commutator.
Substituting the definition of the Jamio\l{}kowski state into the equation above leads to
\begin{equation}
     \mathcal{E}^\dag \star O  = \frac{1}{2}\sum_{ij} \lbrace O,\ketbra{i}{j}\rbrace \otimes \mathcal{E}^\dag (\ketbra{j}{i}) \, .
\end{equation}
This particular form of the QOOT, inspired by the Jordan product state over time in Ref.~\cite{QuantumBayesRetrodiction}, has desirable properties, including linearity in the reference observable, linearity in the quantum map and it is a Hermitian operator.
We can now revisit the necessary and sufficient conditions for the Jordan product QOOT to be well defined, i.e., we have to make sure that the defining properties Eq.~\eqref{minimal_property_1} and Eq.~\eqref{minimal_property_2} are satisfied.
In particular, taking the partial trace of Eq.~\eqref{jordan_product_QOOT_definition} over $\mathcal{H}_B$ leads to
\begin{equation}
\begin{split}
    \Tr_B [\mathcal{E}^\dagger \star O]
    & = \frac{1}{2}\sum_{ij} \lbrace O,\ketbra{i}{j}\rbrace \Tr[\mathcal{E}^\dag (\ketbra{j}{i})] \\
    & = \frac{1}{2}\sum_{ij} \lbrace O,\ketbra{i}{j}\rbrace \bra{i}\mathcal{E}(\mathcal{I})\ket{j} \\
    &= \frac{1}{2}\lbrace O,\mathcal{E}(\mathcal{I})\rbrace \, , 
\end{split}
\end{equation}
where we have used the definition of adjoint map, namely $\Tr[\mathcal{E}^\dag \ketbra{j}{i}] = \Tr [\ketbra{j}{i} \mathcal{E}(\mathcal{I})]$.
For Eq.~\eqref{minimal_property_1} to be satisfied we need 
\begin{equation}
    \lbrace O,\mathcal{E}(\mathcal{I})\rbrace = 2 O
    \label{condition_partial_trace}
\end{equation}
to hold.
Clearly, if $\mathcal{E}$ is unital the condition is satisfied, however that is not the only possibility.
In fact, one easily shows that the necessary and sufficient condition for Eq.~\eqref{condition_partial_trace} to hold is 
\begin{equation}
    \mathcal{E}(\mathcal{I}) = \mathcal{I} + \overline{O} \, ,
\end{equation}
where $\overline{O}$ is a generic operator that anti-commutes with $O$.
Analogously, the partial trace of the Jordan product QOOT over $\mathcal{H}_A$ gives
\begin{equation}
\begin{split}
    \Tr_A [\mathcal{E}^\dag \star O]  & = \frac{1}{2}\sum_{ij} \Tr [\lbrace O,\ketbra{i}{j}\rbrace ] \mathcal{E}^\dag (\ketbra{j}{i}) \\ & = \sum_{ij} \bra{j}O\ket{i} \mathcal{E}^\dag (\ketbra{j}{i}) \\ &= \mathcal{E}^\dag (O) \, ,
\end{split}
\end{equation}
where in the last equality we used the linearity of $\mathcal{E}^\dag$.
Notice that Eq.~\eqref{minimal_property_2} is in this case automatically satisfied. 
We are now ready to substitute the expression of the Jordan product QOOT into Eqs.~\eqref{bayes_rule_pre_processing} and \eqref{bayes_rule_post_processing}, and solve them for $\mathcal{P}$ and $\mathcal{R}$, respectively.

\subsection{Pre-processing protocol}
\label{section_pre_processing}
Let us consider a reference observable $O$ acting on $\mathcal{H}_A$ that we want to protect from a noisy process modeled by a CPTP map $\mathcal{E}: \mathcal{L}(\mathcal{H}_B) \rightarrow  \mathcal{L}(\mathcal{H}_A)$.
As outlined in Sec.~\ref{sec_recovery_maps}, a pre-processing map $\mathcal{P} :\mathcal{L}(\mathcal{H}_A) \rightarrow  \mathcal{L}(\mathcal{H}_B)$ is obtained by solving
\begin{equation}
    \mathcal{E}^\dag \star O = \tau (\mathcal{P}^\dag \star \mathcal{E}^\dag (O)) \, .
\end{equation}
Substituting the expression of the Jordan product QOOT into the previous equation yields
\begin{equation}
    \tau (\lbrace O\otimes\mathcal{I},D[\mathcal{E}^\dag]\rbrace) = \lbrace \mathcal{I} \otimes  \mathcal{E}^\dagger (O),D[{\mathcal{P}^\dag}]\rbrace \, ,
    \label{apply_time_reversal}
\end{equation}
where we have used the fact that applying $\tau$ twice corresponds to the identity channel.
The action of the time-reversal operator $\tau$ on the left-hand side of Eq.~\eqref{apply_time_reversal} reads
\begin{equation}
\begin{split}
     \tau (\lbrace O\otimes\mathcal{I},D[\mathcal{E}^\dag]\rbrace) & = 
     \lbrace \tau(O\otimes\mathcal{I}),\tau(D[\mathcal{E}^\dag)]\rbrace \\ & =
     \lbrace \mathcal{I} \otimes O ,D[\mathcal{E}]\rbrace \, , 
\end{split}
\end{equation}
where we have used the fact that $O$ is Hermitian and that $\tau(D[\mathcal{E}^\dag]) = D[\mathcal{E}]$ (see Appendix \ref{appendix_identities} for a detailed derivation of this identity). 
Hence, we obtain the following equation
\begin{equation}
    \lbrace \mathcal{I} \otimes O , D[\mathcal{E}]\rbrace = \lbrace \mathcal{I} \otimes  \mathcal{E}^\dagger (O),D[{\mathcal{P}^\dag}]\rbrace \, .
    \label{bayes_rule_jordan_product_pre_processing}
\end{equation}
As already mentioned in the previous section, for the two Jordan product QOOTs appearing in the equation above to be well-defined, we need the following two conditions to be satisfied
\begin{equation}
    \lbrace O,\mathcal{E}(\mathcal{I}) \rbrace = 2 O \iff \mathcal{E}(\mathcal{I}) = \mathcal{I} + \overline{O} \, , 
    \label{pre_processing_condition_1}
\end{equation}
\begin{equation}
    \lbrace \mathcal{E}^\dag (O),\mathcal{P}(\mathcal{I}) \rbrace = 2 \mathcal{E}^\dag (O) \iff \mathcal{P}(\mathcal{I}) = \mathcal{I}+\overline{\mathcal{E}^\dag (O)} \, .
     \label{pre_processing_condition_2}
\end{equation}
The condition given in Eq.~\eqref{pre_processing_condition_1} may be directly checked using the input of the problem, namely $O$ and $\mathcal{E}$. 
On the other hand, in what follows we show that Eq.~\eqref{pre_processing_condition_2} is automatically satisfied by any solution $\mathcal{P}$ to Eq.~\eqref{bayes_rule_jordan_product_pre_processing}.
Let us now consider the eigendecomposition of the noisy reference observable, namely 
\begin{equation}
    \mathcal{E}^\dag (O) = \sum_k q_k \ketbra{\omega_k} \, ,
\end{equation}
where for now we assume that the eigenvalues $\lbrace q_k \rbrace$ are such that $q_k + q_j \neq 0 $ for all $j$ and $k$ (in Sec.~\ref{sec_examples} we show how an appropriate regularization of the reference observable may be employed to address the cases where this condition is not satisfied). In Appendix \ref{appendix_pre_processing} we show that the solution to Eq.~\eqref{bayes_rule_jordan_product_pre_processing} reads
\begin{equation}
    \label{recovery_map_pre_processing}
  \mathcal{P}^\dag (\ketbra{\omega_k}{\omega_\ell}) = (q_k + q_\ell)^{-1} \lbrace O,\mathcal{E}(\ketbra{\omega_k}{\omega_\ell}) \rbrace \, .
\end{equation}
This map is clearly HP, and we can easily show that $\mathcal{P}^\dag$ is TP as well (i.e., $\mathcal{P}$ is unital), in fact
\begin{equation}
\begin{split}
     \Tr[ \mathcal{P}^\dag (\ketbra{\omega_k}{\omega_\ell}) ] & =
     (q_k + q_\ell)^{-1} \Tr [\lbrace O,\mathcal{E}(\ketbra{\omega_k}{\omega_\ell}) \rbrace] 
     %\\ & = 2 (q_k + q_\ell)^{-1} \Tr[ O\mathcal{E}(\ketbra{\omega_k}{\omega_\ell})]  
     %\\ & = 2 (q_k + q_\ell)^{-1} \Tr[ \mathcal{E}^\dag(O)\ketbra{\omega_k}{\omega_\ell}] 
     \\ & =  2 (q_k + q_\ell)^{-1} \bra{\omega_\ell}\mathcal{E}^\dag(O)\ket{\omega_k} \\ & = 2 (q_k + q_\ell)^{-1} q_k \delta_{k\ell} = \delta_{k\ell} \, ,
\end{split}
\end{equation}
where $\delta_{k\ell}$ denotes the Kronecker delta.
A consequence of the fact that $\mathcal{P}^\dag$ is HPTP is that can be expressed as a linear combination of CPTP maps.
Additionally, in Appendix \ref{appendix_invertible_channel_recovery} we prove the desirable property that 
if the map $\mathcal{E}$ corresponds to a unitary evolution (i.e., it is invertible and its inverse is CPTP), then $\mathcal{E}^{-1}$ is indeed the pre-processing map $\mathcal{P}$.

\subsection{Post-processing protocol}
As previously outlined in Sec.~\ref{sec_recovery_maps}, a recovery map $\mathcal{R}:\mathcal{L}(\mathcal{H}_\mathcal{B})\rightarrow \mathcal{L}(\mathcal{H}_\mathcal{A})$ for the post-processing protocol is obtained by solving 
\begin{equation}
    \mathcal{R}^\dag \star O = \tau (\mathcal{E}^\dag \star \mathcal{R}^\dag (O)) \, .
\end{equation}
Substituting the expression of the Jordan product QOOT into the equation above yields
\begin{equation}
      \lbrace O \otimes \mathcal{I},  D[\mathcal{R}^\dag]\rbrace =  \tau ( \lbrace \mathcal{I} \otimes \mathcal{R}^\dag(O), D[\mathcal{E}^\dag] \rbrace ) \, .
    \label{bayes_rule_jordan_product_post_processing_1}
\end{equation}
Analogously to the pre-processing protocol, for the two Jordan product QOOTs appearing in Eq.~\eqref{bayes_rule_jordan_product_post_processing_1} to be well-defined we need the following conditions to hold
\begin{equation}
    \lbrace O,\mathcal{R}(\mathcal{I}) \rbrace = 2 O \iff \mathcal{R}(\mathcal{I}) = \mathcal{I} + \overline{O} \, , 
    \label{post_processing_condition_1}
\end{equation}
\begin{equation}
    \lbrace \mathcal{R}^\dag (O),\mathcal{E}(\mathcal{I}) \rbrace = 2 \mathcal{R}^\dag (O) \iff \mathcal{E}(\mathcal{I}) = \mathcal{I}+\overline{\mathcal{R}^\dag (O)} \, .
     \label{post_processing_condition_2}
\end{equation}
Coming back to Eq.~\eqref{bayes_rule_jordan_product_post_processing_1}, the action of the time-reversal operator $\tau$ reads
\begin{equation}
\begin{split}
         \tau ( \lbrace \mathcal{I} \otimes \mathcal{R}^\dag(O), D[\mathcal{E}^\dag] \rbrace ) & = \lbrace \tau(\mathcal{I} \otimes \mathcal{R}^\dag (O) ), \tau ( D[\mathcal{E}^\dag] ) \rbrace \\ & = 
     \lbrace \mathcal{R}^\dag(O)  \otimes \mathcal{I} , D[\mathcal{E}] \rbrace \, ,
\end{split}
\end{equation}
where we have used the fact that we look for solutions $\mathcal{R}$ to the equation that are HP,  
%\GB{\textbf{(does it need to be enforced? is it satisfied in our examples? check!)}},
%\hk{(HK: I think Eq.~(25) should hold even with a weaker condition that ${\cal R}^\dagger(O)$ is Hermitian for a given $O$. Furthermore, I think this can be shown from Eq.~(25) that the LHS, ${\cal R}^\dagger \star O$, is Hermitian (I realzied that this is not the case), so the LHS should also be a Hermitian operator. If ${\cal R}^\dagger(O)$ is non-Hermitian, i.e., having an expression ${\cal R}^\dagger(O) = A + i B$ with Hermitian operators $A$ and $B$, then $\tau\left({\cal E}^\dagger \star {\cal R}^\dagger(O) \right)$ cannot be Hermitian. Does it make sense?)}
$O$ is Hermitian and $\tau (D[\mathcal{E}^\dag]) = D[\mathcal{E}]$.
Hence, we obtain
\begin{equation}
         \label{bayes_rule_jordan_product_post_processing_2}
        \lbrace O\otimes\mathcal{I},D[\mathcal{R}^\dag]\rbrace =  \lbrace  \mathcal{R}^\dag (O)\otimes\mathcal{I}  , D[\mathcal{E}]\rbrace \, .
\end{equation}
The difficulty in solving this equation lies in the fact that, unlike the pre-processing protocol, the recovery map now appears on both sides of the equation and, more importantly, we do not know anything about $\mathcal{R}^\dag(O)$ a priori. In the following, we briefly outline the techniques used to solve Eq.~\eqref{bayes_rule_jordan_product_post_processing_2}, while a detailed derivation can be found in Appendix \ref{appendix_qubit_system}.
First, let us consider the eigen-decomposition of the reference observable, i.e. $O=\sum_i q_i \ketbra{\omega_i}$.
Notice that, while we do not know what $\mathcal{R}^\dag(O)$ is, we may express it as $\mathcal{R}^\dag(O)=\sum_i q_i \mathcal{R}^\dag(\ketbra{\omega_i})$, by linearity.
Projecting both sides of Eq.~\eqref{bayes_rule_jordan_product_post_processing_2} onto the joint Hilbert space basis $\lbrace \ket{\omega_i}\otimes\ket{\omega_j}\rbrace$ yields
%\begin{equation}
    %(q_k + q_\ell) \mathcal{R}^\dag(\ketbra{\omega_\ell}{\omega_k})  =  \lbrace \mathcal{R}^\dag(O),\mathcal{E}(\ketbra{\omega_\ell}{\omega_k}) \rbrace \, .
%\end{equation}
\begin{equation}
    (q_k + q_\ell) \mathcal{R}^\dag(\ketbra{\omega_\ell}{\omega_k})  =  \sum_i q_i \lbrace \mathcal{R}^\dag(\ketbra{\omega_i}),\mathcal{E}(\ketbra{\omega_\ell}{\omega_k}) \rbrace \, .
\end{equation}
The equation above can be recast as
\begin{align}
    X_{k\ell} = \sum_i \lbrace X_{ii},A_{k\ell} \rbrace \, ,
    \label{post_processing_recasted_main_text}
\end{align}
where $X_{k\ell} = (q_k + q_\ell)  \mathcal{R}^\dag (\ketbra{\omega_k}{\omega_\ell})$, and $A_{k\ell}  = \mathcal{E}(\ketbra{\omega_k}{\omega_\ell})/2$.
The HP requirement for the recovery map is equivalent to enforcing $X_{k\ell}^\dag = X_{\ell k}$.
Lastly, we need to make sure that the conditions necessary and sufficient for the QOOTs to be well defined are satisfied.
In particular, Eq.~\eqref{post_processing_condition_2} can be recast as $0 = \sum_i \lbrace X_{ii},\mathcal{E}(\mathcal{I})-\mathcal{I} \rbrace$. A unital noise map always satisfies this constraint, however in general this would impose an additional constraint on the recovery map $\mathcal{R}$.
Similarly, one can show (see Appendix \ref{appendix_trace_contraint}) that Eq.~\eqref{post_processing_condition_1} is equivalent to imposing the additional constraint $\Tr[X_{k\ell}] = q_k \delta_{k\ell}$. Notice how this is different from the pre-processing scenario, where the arising recovery maps $\mathcal{P}$ are automatically unital and hence satisfy Eq.~\eqref{pre_processing_condition_2} without needing to impose additional constraints.

Let us now consider the case of a qubit system, meaning that we can decompose all operators appearing in the equation in the Pauli basis, namely
\begin{align}
    X_{k\ell} = \sum_{j=0}^3 x^{(k\ell)}_j \sigma_j \, , \quad A_{k\ell} = \sum_{j=0}^3 a^{(k\ell)}_j \sigma_j \, ,
\end{align}
where $\lbrace \sigma_0,\sigma_1,\sigma_2,\sigma_3 \rbrace \equiv \lbrace\mathcal{I},X,Y,Z \rbrace$ denote the Pauli operators.
We can now substitute these expressions into Eq.~\eqref{post_processing_recasted_main_text} and obtain an equivalent system of linear equations for the eight coefficients $x_j^{(k\ell)}$ 
\begin{align}\label{lineqns}
    x^{(kl)}_0 & = 2 \sum_{i,j} x^{(ii)}_j a^{(kl)}_j \, , \\ 
    x^{(kl)}_j & = 2 \sum_{i} \left( x^{(ii)}_0 a^{(kl)}_j  +  x^{(ii)}_j a^{(kl)}_0  \right) \, , \quad j\in\lbrace 1,2,3 \rbrace \, .
    \label{equivalent_equation1}
\end{align}
In Appendix \ref{appendix_d_dimensional_system} we outline a detailed derivation of the system of equation above and a generalization to $d-$dimensional systems.
Analogously to the pre-processing case, if $\mathcal{E}$ is a unitary evolution (i.e., the inverse exists and it is CPTP) then $\mathcal{E}^{-1}$ is a recovery map $\mathcal{R}$ that solves Eq.~\eqref{bayes_rule_jordan_product_post_processing_2}, as can be shown by direct substitution.

\section{Simulation of recovery maps and application to error mitigation}
\label{sec_simulating_recovery_maps}
Physically-implementable operations are described using CPTP maps. More generally, if we allow these operations to have a probabilistic rather than deterministic implementation, they can be characterized by completely-positive, trace non-increasing (CPTN) maps.
Nevertheless, non-physical maps find application in several quantum information processing tasks.
In fact, while an HP $-$ but non-CP $-$ map $\mathcal{R}$ represents a physically-unattainable quantum operation, it is always possible to express it as a linear combination of CPTN maps $\lbrace \mathcal{F}_i \rbrace$, namely
\begin{equation}
    \mathcal{R} = \sum_i c_i \mathcal{F}_i \, ,
    \label{QPD}
\end{equation}
where $c_i$ are real (possibly negative) coefficients.
In particular, we might think at the total negativity as a measure of the ``unphysicality'' of the map.
We will refer to Eq.~\eqref{QPD} as a quasi-probability decomposition (QPD) of $\mathcal{R}$.
If $\mathcal{R}$ is also TP, it turns out that we can always decompose it in terms of CPTP maps and that $\sum_i c_i = 1$.
A QPD of an unphysical map $\mathcal{R}$ may be used as a simulation tool to estimate expectation values of the form $\Tr [\mathcal{R}(\rho) O]$.
In fact, we may express the latter as 
\begin{equation}
\begin{split}
    \Tr [\mathcal{R}(\rho) O] & = \sum_i c_i \Tr [\mathcal{F}_i(\rho) O] \\ & = \gamma \sum_i \textrm{sgn}(c_i) \, p_i \Tr [\mathcal{F}_i(\rho) O] \, ,
\end{split}
\end{equation}
where $\gamma = \sum_i \vert c_i \vert$ and $p_i = \vert c_i \vert / \gamma$ is a probability distribution.
The idea is now to sample a quantum channel $\mathcal{F}^{(j)}$ from $\lbrace \mathcal{F}_i \rbrace$ according to the probability distribution $p_i$, apply it to the state $\rho$ and subsequently measure the observable $O$ and record the measurement outcome $o^{(j)}$ ($j$ labels the sample number). After $N$ sampling rounds, we can construct the following unbiased estimator for $\Tr [\mathcal{R}(\rho) O]$
\begin{equation}
    E = \frac{\gamma}{N}\sum_{j=1}^N \textrm{sgn} \,(c^{(j)}) o^{(j)} \, .
    \label{estimator_bias_free}
\end{equation}
Furthermore, from Hoeffding's inequality \cite{hoeffding1994probability} the number of samples required for $E$ to be $\epsilon-$close to the true value $\Tr[\mathcal{R}(\rho) O]$, with probability no less than $1-\delta$, scales as $N\propto \gamma^2 \log(2/\delta)/\epsilon^2$ \cite{zhao2023power}.
The parameter $\gamma$ can then be used as a measure of the \emph{simulation cost} of $\mathcal{R}$, according to the decomposition Eq.~\eqref{QPD}.
It is clear that different decompositions of the same unphysical map $\mathcal{R}$ lead to different $\gamma$ and different simulations costs as a result. In Ref.~\cite{physical_implementability_linear_maps}, the authors show that the optimal decomposition (i.e., achieving the minimum value of $\gamma$) can be computed via semi-definite programming.

A prominent example of the simulation technique described above is provided by probabilistic error cancellation \cite{temme_error_mitigation}, an error mitigation technique aimed at producing bias-free expectation values for systems that are subject to a known noise model, characterized by the CPTP map $\mathcal{E}$. Assuming that the noise map can be mathematically inverted, the key insight behind PEC is that we can simulate $\mathcal{E}^{-1} = \sum_i c_i \mathcal{F}_i$ (in the sense discussed above, since the inverse of a CPTP map is HPTP) and obtain bias-free estimates of the noiseless expectation value $\Tr [\rho O]$ by performing measurements of $O$ on $\mathcal{F}_i (\mathcal{E}(\rho))$ and by  classical post-processing of data. Within this framework, the simulation cost we mentioned earlier can also be interpreted as \emph{sampling overhead}, as it can be showed \cite{temme_error_mitigation,review_error_mitigation} that the cost associated with obtaining a bias-free estimate of the noiseless expectation value is an increase in the variance of the estimator. In particular, the ratio between the variance of the bias-free estimator and that of the noisy estimator can be shown to scale with $\gamma^2$.

Inspired by PEC, our recovery maps may also be employed to obtain noise-free expectation values of a quantum observable $O$ subject to a CPTP noise map $\mathcal{E}$.
In fact, by construction we have that
\begin{equation}
    \Tr[\rho O] = \Tr[\rho \, \mathcal{E}^\dag (\mathcal{R}^\dag (O))] \, ,
\end{equation}
for the post-processing protocol, and 
\begin{equation}
    \Tr[\rho O] = \Tr[\rho \, \mathcal{P}^\dag (\mathcal{E}^\dag (O))] \, ,
\end{equation}
for the pre-processing protocol.
All we need to do, is compute the appropriate map ($\mathcal{R}$ or $\mathcal{P}$) as described in Section \ref{sec_jordan_product}, and decompose it in terms of physically-implementable maps.
Our protocols might prove particularly useful when the inverse of the noise channel does not exist, hence the framework of PEC is not applicable.
Furthermore, our protocols would also be useful in situations where the implementation cost of our recovery maps outperform that of the inverse noise map.

%\section{Recovery protocols}
%Generally, HP but non-CPTP maps represent physically unattainable quantum operations. Nonetheless, the action of a HP map can be simulated by the action of some set of completely-positive, trace non-increasing (CPTN) maps $\{\mathcal{F}_i\}$ according to a linear decomposition of the HP map in terms of the CPTN maps
%\begin{equation}\label{decomposition}
%   \mathcal{R} = \sum_{i} \alpha_i \mathcal{F}_i \, .
%\end{equation}
%The simulation of non-physical operations has previous seen application in probabilistic error cancellation (PEC), an error mitigation technique for bias-free estimation of expectation values $\langle O \rangle$ from noisy data $\mathcal{E}(\rho)$ through simulation of the inverse noise channel $\mathcal{E}^{-1}$. Inheriting this intuition, the action of the recovery maps $\mathcal{R}$ derived from Eqs. \eqref{bayes_rule_pre_processing} and \eqref{bayes_rule_post_processing} can be actualised through sampling and probabilistic application of the physical CPTN maps, before classical post-processing of the noisy data by weighting expectation values according to the coefficients $\alpha_i$ in Eq. \eqref{decomposition}.

%\subsection{Pre-processing protocol}
%First, the case of pre-processing recovery operations is considered, where the ordering of operations in the Heisenberg picture are as described by Eq. \eqref{pre_processing_recovery_property}. In this case, the solution of Eq. \eqref{bayes_rule_pre_processing} following in exact analogy

\section{Examples}
\label{sec_examples}
In this Section, we provide explicit constructions of recovery maps (pre- and post-processing protocols) for Pauli reference observables and physically-relevant noise models, namely generalized amplitude damping (GAD) and stochastic Pauli noise.
The latter is of particular importance, since Pauli twirling \cite{Pauli_twirling} may be employed to convert an arbitrary noise channel into a stochastic Pauli noise channel.
For all examples, we provide a decomposition of the recovery maps in terms of physically-implementable channels and compute the cost of implementation $\gamma$.
Comparing the latter with known optimality results reported in Ref.~\cite{information_recoverability} reveals that the recovery maps here presented and their decompositions are optimal.

\subsection{GAD channel and Pauli X (pre-processing)}\label{subsec:ExamplesGADX}
The generalized amplitude damping channel can be viewed as the qubit analogue of the bosonic thermal channel \cite{GADWilde}, and may be expressed as $\mathcal{E}(\rho) = \sum_{j=1}^4 K_j \rho K_j^\dag$, where the Kraus operators read
\begin{align}
 & K_1 = \sqrt{p}\ketbra{0} + \sqrt{p(1-\epsilon)}\ketbra{1} \, , \\ 
 & K_2 = \sqrt{p\epsilon} \ketbra{0}{1} \, , \\ 
 & K_3 = \sqrt{(1-p)(1-\epsilon)} \ketbra{0} + \sqrt{(1-p)}\ketbra{1} \, , \\ 
 &  K_4 = \sqrt{\epsilon(1-p)}\ketbra{1}{0} \, ,
\end{align}
with $0\leq p\leq 1$ and $0 \leq \epsilon\leq 1$. 
One easily shows that the GAD channel is non-trivially unital only for $p=1/2$ and that the amplitude damping channel can be retrieved by setting $p=1$.
In this example, we set the reference observable $O$ to be the Pauli $X$ operator, and look for a pre-processing map $\mathcal{P}$, i.e. satisfying $\mathcal{P}^\dag (\mathcal{E}^\dag (X)) = X$.
Notice that the necessary and sufficient  condition in Eq.~\eqref{pre_processing_condition_1} for the quantum observable over time (hence, the task) to be well defined is satisfied, since $\mathcal{E}(\mathcal{I}) = \mathcal{I}+\epsilon (2p-1)Z$ and $\lbrace X,Z \rbrace = 0$.
In Section \ref{section_pre_processing} we showed that the expression for such a recovery map reads
\begin{equation}
  \mathcal{P}^\dag (\ketbra{\omega_k}{\omega_\ell}) = (q_k + q_\ell)^{-1} \lbrace O,\mathcal{E}(\ketbra{\omega_k}{\omega_\ell}) \rbrace \, ,
  \label{recovery_map_expression_pre_processing}
\end{equation}
where $q_k$ and $\ket{\omega_k}$, are the eigenvalues and eigenstates of the noisy reference observable $\mathcal{E}^\dag (O)$, respectively.
We remind the reader that the expression above is formally valid for $q_k + q_\ell \neq 0$.
However, this condition condition is not satisfied in our scenario, since $\mathcal{E}^\dag (X) = \sqrt{1-\epsilon} X$ and $X$ is traceless.
We may work around this potential setback by \emph{regularizing} the reference observable, i.e. we define 
\begin{equation}
    \tilde{X} = 
    \begin{pmatrix}
        \lambda & 1 \\ 1 & \lambda 
    \end{pmatrix} = \lambda\mathcal{I}+X\, ,
\end{equation}
where $\lambda$ is a positive regularization parameter that will be taken the limit to zero at the end of the calculations.
The action of the (adjoint) GAD channel on $\tilde{X}$ reads 
\begin{equation}
    \mathcal{E}^\dag (\tilde{X}) =  \lambda \mathcal{I} + \sqrt{1-\epsilon} X = q_1 \ketbra{\omega_1} + q_2 \ketbra{\omega_2} \, ,
\end{equation}
where
\begin{align}
      & q_1 = \lambda - \sqrt{1-\epsilon}\, , \quad  \ket{\omega_1} = \frac{-\ket{0} + \ket{1}}{\sqrt{2}} \equiv -\ket{-}\, , \\
      & q_2 = \lambda + \sqrt{1-\epsilon} \, , \quad \ket{\omega_2} = \frac{\ket{0} + \ket{1}}{\sqrt{2}} \equiv \ket{+} \, . 
\end{align}
We can now compute the pre-processing map via Eq.~\eqref{recovery_map_expression_pre_processing} and, after some algebra, obtain
%\begin{align}
%    \mathcal{R}^\dag (\ketbra{-}) &  = \frac{1}{2}\mathcal{I} -\frac{1}{2\sqrt{1-\epsilon}} X \, ,\\ \mathcal{R}^\dag (\ketbra{-}{+}) &  = \frac{1-\epsilon}{2} Z +i\frac{\sqrt{1-\epsilon}}{2} Y \, , \\ \mathcal{R}^\dag (\ketbra{+}{-}) &  = \frac{1-\epsilon}{2} Z - i \frac{\sqrt{1-\epsilon}}{2} Y \, , \\ \mathcal{R}^\dag (\ketbra{+}{+}) &  = \frac{1}{2}\mathcal{I} +\frac{1}{2\sqrt{1-\epsilon}} X  \, ,
%\end{align}
\begin{align}
    \mathcal{P}^\dag (\mathcal{I}) &  = \mathcal{I} \, , \\ 
    \mathcal{P}^\dag (X) &  = \frac{X}{\sqrt{1-\epsilon}} \, , \\ 
    \mathcal{P}^\dag (Y) &  =  \sqrt{1-\epsilon} Y \, , \\ 
    \mathcal{P}^\dag (Z) &  =  (1-\epsilon) Z\, ,
\end{align}
where we have already taken the limit $\lambda \rightarrow 0$ (the potentially problematic $1/\lambda$ dependency of the terms $\mathcal{P}^\dag (\ketbra{+}{-})$ and $\mathcal{P}^\dag (\ketbra{-}{+})$ actually cancels out with a $\lambda$ factor coming out of the numerator). 
One can easily check that this map indeed satisfies the recovery property, i.e. $\mathcal{P}^\dag (\mathcal{E}^\dag (X)) = X$.
Interestingly, we note that although $\mathcal{E}^\dag$ and $\mathcal{P}^\dag$ do not commute in general, it is true that 
\begin{equation}
    \label{pre_and_post}
    X = \mathcal{P}^\dag (\mathcal{E}^\dag (X)) = \mathcal{E}^\dag (\mathcal{P}^\dag (X)) \, . 
\end{equation}
Hence, it follows that the recovery map above may also be employed in a post-processing protocol.
 
The next step is to find an operator-sum (i.e., Kraus-like) representation of the recovery map we just found.
In particular, we want to express $\mathcal{P}^\dag$ as a linear combination of physically realizable CP channels.
A common technique to achieve a Kraus-like decomposition of $\mathcal{P}^\dag$, consists in the diagonalization the recovery map's Choi matrix, namely 
\begin{equation}
    \mathcal{J}[\mathcal{P}^\dag] = \sum_{ij}\mathcal{P}^\dag (\ketbra{i}{j})\otimes \ketbra{i}{j} \, ,
\end{equation}
followed by the unvectorization of its eigenvectors.
After some algebra, we obtain the following operator-sum decomposition
\begin{equation}
    \mathcal{P}^\dag(\sigma) = \sum_{i=1}^4 e_i E_i \sigma E_i^\dagger \, ,
    \label{QOOT_example_decomposition}
\end{equation}
where $\lbrace e_i \rbrace$ are the Choi matrix eigenvalues, namely
\begin{equation}
    e_1 = (2-\epsilon)\alpha_+  \, , \,  e_2 = (2-\epsilon) \alpha_- \, , \, e_3 = \epsilon \alpha_+ \, , \,  e_4 = \epsilon \alpha_- \, ,
\end{equation}
with $\alpha_\pm = \frac{1}{2} \pm \frac{1}{2\sqrt{1-\epsilon}}$.
Notice that $e_2$ and $e_4$ are negative, meaning that the Choi matrix of the recovery map is not positive semi-definite, hence the recovery map itself is not CP. 
The operators $E_i$ are obtained from the unvectorization of the Choi matrix (normalized) eigenvectors, in particular we find 
\begin{align}
    E_1 & =\frac{\mathcal{I}}{\sqrt{2}} \, , \,
    E_2 =\frac{Z}{\sqrt{2}}\, , \,
    E_3 = -\frac{X}{\sqrt{2}} \, , \,
    E_4 = \frac{iY}{\sqrt{2}} \, .
\end{align}
Hence, the recovery map can be decomposed in terms of CPTP maps and, in particular, takes the form of a stochastic Pauli map with some negative coefficients.
As previously explained, the cost of implementation of an HPTP map, given a specific linear decomposition in terms of CPTP maps, may be evaluated by summing the absolute value of said coefficients. 
For the above operator-sum representation of $\mathcal{P}^\dag$, the cost of implementation equals to 
\begin{equation}
    \gamma_{\mathcal{P}^\dag} =\frac{1}{2}(\vert e_1 \vert +  \vert e_2 \vert + \vert e_3\vert  +\vert e_4 \vert) = \frac{1}{\sqrt{1-\epsilon}} \, .
    \label{optimal_cost_GAD}
\end{equation}
We may first compare this with the \emph{optimal} (i.e., minimized over all possible decompositions in terms of linear combinations of CPTP maps) implementation cost of the GAD channel inverse map.
In Ref.~\cite{physical_implementability_linear_maps} the authors show that the latter reads
\begin{equation}
    \gamma_{\mathcal{E}^{-1}} = \frac{\vert1-2p\vert \epsilon +1}{1-\epsilon} \, .
\end{equation}
One then shows that $\gamma_{\mathcal{P}^\dag} < \gamma_{\mathcal{E}^{-1}}$ for all possible values of $p$ and $\epsilon$, meaning that the sampling overhead associated with our recovery map with the decomposition above outperforms the traditional PEC protocol relying on implementing the channel inverse.
\\
For a given noise channel and reference observable, the authors of Ref.~\cite{information_recoverability} employed semi-definite programming to optimize the implementation cost across all possible Hermitian-preserving trace-scaling (HPTS) recovery maps, and across their decompositions in terms of CPTP maps. In particular, for the GAD channel and Pauli $X$ reference observable, they demonstrated that the implementation cost is lower bounded by Eq.~\eqref{optimal_cost_GAD}, thus making the recovery map we found \emph{optimal} for this task.

\subsection{Stochastic Pauli channel and Pauli $Z$ (post-processing)}

Another key example is the stochastic Pauli channel, defined as:
\begin{equation}
    \mathcal{E}(\rho) = p_0\rho  + p_1X\rho X+p_2Y\rho Y+p_3 Z\rho Z,
    \label{stochastic_pauli_noise_def}
\end{equation}
where the coefficients $p_i$ constitute a probability distribution, satisfying $p_i \geq 0$ and $\sum_i p_i = 1$.
Eq.~\eqref{stochastic_pauli_noise_def} encompasses several widely employed noise models. For instance, setting $p_1 = p_2 = p_3 = \frac{\lambda}{4}$ and $p_0 = 1-\frac{3\lambda}{4}$ yields the depolarizing channel, while choosing $p_0 = 1-p, \ p_1=p_2=0, \ p_3=p$ for $0< p < 1$ results in the dephasing channel.  
Furthermore, the stochastic Pauli channel is of particular interest because an arbitrary single-qubit noise channel may be reduced to a stochastic Pauli noise channel through Pauli twirling \cite{Pauli_twirling}.
In this example, we set the reference observable $O$ to be the Pauli $Z$ operator (analogous results would follow for $X$ or $Y$) and seek a post-processing recovery map $\mathcal{R}$, hence satisfying $\mathcal{E}^\dagger (\mathcal{R}^\dagger (Z)) = Z$.
Since any stochastic Pauli channel is unital, the consistency condition in Eq.~\eqref{post_processing_condition_2} is automatically satisfied. On the other hand, the remaining condition in Eq.~\eqref{post_processing_condition_1} still needs to be imposed.
We have previously demonstrated that a post-processing recovery map can be obtained by solving the following equation
\begin{equation}
    (q_k + q_\ell) \mathcal{R}^\dag(\ketbra{\omega_\ell}{\omega_k})  =  \sum_i q_i\lbrace \mathcal{R}^\dag(\ketbra{\omega_i}),\mathcal{E}(\ketbra{\omega_\ell}{\omega_k}) \rbrace , 
\end{equation}
where $q_k$ and $\ket{w_k}$ are the eigenvalues and corresponding eigenstates of the reference observable, respectively.
Similar to the previous example, we regularize the reference observable by defining $\tilde{Z} = Z + \lambda \mathcal{I}$, where $\lambda \geq 0$ serves as a real regularization parameter that will be taken to zero at the end of the calculation. 
The eigendecomposition of $\tilde{Z}$ is thus characterized by
\begin{align}
    q_1 &= 1 + \lambda \, , \quad \ket{\omega_1} = \ket{0},\\
    q_2 &= -1+\lambda \, , \quad \ket{\omega_2} = \ket{1}.
\end{align}
As shown in Sec.~\ref{sec_jordan_product}, computing the recovery map amounts to solving the system of linear equations given by Eq.~\eqref{lineqns} and Eq.~\eqref{equivalent_equation1} with the coefficients $a_j^{(k \ell)}$ reading
\begin{equation}
    a_0^{(k\ell)} = \frac{1}{2}\omega_0^{(k\ell)}\, , \quad a_j^{(k\ell)} = \left(p_0 + p_j - \frac{1}{2}\right)\omega_j^{(k\ell)} \, .
\end{equation}
Here, $\omega_j^{(k\ell)}$ are the Pauli basis coefficients for the outer product of the basis elements, namely
\begin{equation}
    \ketbra{\omega_k}{\omega_\ell} = \sum_{j} \omega_j^{(k\ell)}\sigma_j \, .
\end{equation}
The solution to these equations, upon taking $\lambda \to 0$, yields the following recovery map:
\begin{equation}
\begin{split}
    & \mathcal{R}^{\dagger}(\mathcal{I}) = \mathcal{I}\, , \\ &
    \mathcal{R}^{\dagger}(X) = (p_0+p_1-p_2-p_3)X\, , \\ & 
    \mathcal{R}^{\dagger}(Y) = (p_0-p_1+p_2-p_3)Y\, ,\\ & 
     \mathcal{R}^{\dagger}(Z) = \frac{1}{p_0-p_1-p_2+p_3}Z\, .
     \end{split}
\end{equation}
The recovery map $\mathcal{R}$ is thus well defined for all stochastic Pauli channels except for the particular cases of $p_0 + p_3 = \frac{1}{2}$. Furthermore, since $\mathcal{R}^\dag$ is manifestly TP, it implies that $\mathcal{R}$ is unital, thereby satisfying the consistency condition in Eq.~\eqref{post_processing_condition_1}.
Next, we want to express the recovery map as a linear combination of physically realizable maps.
To this end, following the procedure outlined in the previous example, we compute the Choi matrix of $\mathcal{R}^\dag$, diagonalize it, and unvectorize its (normalized) eigenvectors. After some algebra, we obtain the following operator-sum representation
\begin{align}
    \label{QPD_decomposition_SPC_recovery}
    \mathcal{R}^{\dagger}_Z(\sigma) = \ & \frac{\alpha_0+\beta_0}{2}\sigma + \frac{\alpha_1+\beta_1}{2}X\sigma X\nonumber\\
    &+ \frac{\alpha_1-\beta_1}{2}Y\sigma Y + \frac{\alpha_0-\beta_0}{2}Z\sigma Z \, .
\end{align}
%\begin{equation}
%\renewcommand{\arraystretch}{0.8}
    %\mathcal{J}[\mathcal{R}^{\dagger}] = \begin{pmatrix}
        %\alpha_0 & 0 & 0 &\beta_0 \\
        %0 & \alpha_1 & \beta_1 & 0 \\
        %0 & \beta_1 & \alpha_1 & 0 \\
        %\beta_0 & 0 & 0 & \alpha_0
    %\end{pmatrix}\, ,
%\renewcommand{\arraystretch}{1.0} 
%\end{equation}
where
\begin{align}
    \alpha_0 &= \frac{p_0+p_3}{p_0-p_1-p_2+p_3}\, , \quad \beta_0 = p_0-p_3\, ,\nonumber\\
    \alpha_1 &= -\frac{p_1+p_2}{p_0-p_1-p_2+p_3}\, , \quad \beta_1  = p_1-p_2\, .
\end{align}
Therefore, we found that the recovery map is also a stochastic Pauli map, albeit with some negative coefficients, which render it non-CP.
Finally, we sum the modulus of the expansion's coefficients to compute the cost of implementation of $\mathcal{R}$, according to the quasi-probability decomposition in Eq.~\eqref{QPD_decomposition_SPC_recovery}. 
After some algebra, we obtain 
\begin{equation}
   \gamma_{\mathcal{R}^{\dagger}} = \frac{1}{|p_0-p_1-p_2+p_3|} \, .
\end{equation}
Similarly to the previous example, this cost is optimal \cite{information_recoverability}, meaning that no non-CPTP operation with the associated recovery property can be simulated by the quasi-probability decomposition method with a smaller sampling overhead.
Analogous recovery maps, along with their corresponding sampling costs and optimality properties, can be derived for the other Pauli observables.

\section{Conclusions}
\label{sec_conclusions}
In this work, we introduced the concept of QOOT as a dual object of the QSOT, allowing us to jointly describe two observables that are time-like separated, while also capturing their temporal correlations. The main difference compared with QSOT is that the construction of the QOOT is not universally applicable. We established a no-go theorem for constructing QOOT by fully characterizing the condition on the pairs of prior observables and noise maps to properly define a QOOT.
We then leveraged QOOTs to systematically define recovery maps via an equation that reflects a certain time-reversal symmetry of the system, establishing a notion of time-reversal for generic quantum channels, with respect to a reference observable.
By construction, these time-reversal maps act as recovery maps for the reference observable and, although generally non-CP, they can be decomposed into physically-realizable maps, enabling their application in error mitigation tasks.
We introduced a specific instance of the QOOT, based on the Jordan product, and used it to provide explicit examples of recovery maps for widely employed noise models. 
For the noise models we studied, our approach achieves optimal performance, measured in terms of sampling overhead, outperforming the traditional probabilistic error cancellation protocol.

This work leaves interesting open questions that are worth exploring in future research. First, it would be valuable to investigate how general the optimality result is, and to further understand the underlying reasons for its validity. Additionally, while we demonstrated that the QOOT, as defined here, may not always be constructible when the noise channel is non-unital, it remains an open question whether an alternative framework can be developed to ensure well-defined recovery maps under fully general conditions.
It would also be interesting to explore extending the formalism to continuous variable systems, and to compare the performance of our error mitigation protocol with that of the PEC approach  \cite{undoing_photon_loss}.

\section{Acknowledgments}
G.B. is part of the AppQInfo MSCA ITN which received funding from the European Union’s Horizon 2020 research and innovation programme under the Marie Sklodowska-Curie grant agreement No 956071.
H.K. is supported by KIAS individual grant (CG085301) at the Korea Institute for Advanced Study.
This work was also supported by the National Research Foundation of Korea (NRF) grant funded by the Korea government (MSIT) (No. RS-2024-00413957).
M.S.K. acknowledged support from UK EPSRC EP/W032643/1, EP/Y004752/1 and EP/Z53318X/1.

\bibliography{biblio}

\onecolumngrid

\appendix

\section{Useful identities}
\label{appendix_identities}
In this Appendix we prove that, for a   CPTP map $\mathcal{E}$ it holds true that $\tau(D[\mathcal{E}^\dag])= D[\mathcal{E}]$.
\begin{equation}
\begin{split}
    D[\mathcal{E}] & = \sum_{ij} \ketbra{i}{j} \otimes \mathcal{E}(\ketbra{j}{i}) = \sum_{ij\ell} \ketbra{i}{j} \otimes K_\ell \ketbra{j}{i} K_\ell^\dag = \sum_{ij\ell n m} \ketbra{i}{j}  \otimes \ketbra{n} K_\ell \ketbra{j}{i} K_\ell^\dag \ketbra{m} \\ 
     & =  \sum_{ij\ell n m} \ketbra{i} K_\ell^\dag \ketbra{m}{n}K_\ell \ketbra{j} \otimes \ketbra{n}{m} = \sum_{mn\ell} K_\ell^\dagger \ketbra{m}{n} K_\ell \otimes \ketbra{n}{m} = \sum_{nm} \mathcal{E}^\dag (\ketbra{m}{n}) \otimes \ketbra{n}{m} 
     %= \gamma(D[\mathcal{E}^\dag] 
     \, ,
\end{split}
\end{equation}
where we have used the Kraus decomposition of $\mathcal{E}$, namely $\mathcal{E}(\rho) = \sum_\ell K_\ell \rho K^\dag_\ell$.
Hence, we obtain $\tau (D[\mathcal{E}^\dag]) = (D[\mathcal{E}])^\dag = D[\mathcal{E}]$, where the last equality follows from $\mathcal{E}$ being CP.
%\GB{($\gamma$ has not been introduced before in the main text, yet )}

\section{Pre-processing protocol recovery map}
\label{appendix_pre_processing}
In this Appendix we obtain the pre-processing map, by solving the following equation 
\begin{equation}
    \lbrace \mathcal{I} \otimes O,D[\mathcal{E}] \rbrace = \lbrace \mathcal{I}\otimes\mathcal{E}^\dag (O),D[\mathcal{P}^\dag]\rbrace \, .
    \label{eq_appendix_pre_processing}
\end{equation}
The reader may use Figure \ref{fig_pre_processing} as a reference for the spaces that maps and operators act upon.
Let us consider the eigen-decomposition of the noisy reference observable, i.e. $\mathcal{E}^\dag (O)=\sum_k q_k \ketbra{\omega_k}$, and project both sides of Eq.~\eqref{eq_appendix_pre_processing} onto the joint Hilbert space $\mathcal{H}_A \otimes \mathcal{H}_B$ basis $\lbrace \ket{\omega_i}\otimes\ket{\omega_j} \rbrace \equiv \lbrace \ket{\omega_i}\ket{\omega_j} \rbrace$. 
In particular, the l.h.s. reads
\begin{equation}
\begin{split}
    \bra{\omega_i} \bra{\omega_k} \lbrace \mathcal{I} \otimes O ,D[\mathcal{E}] \rbrace \ket{\omega_j} \ket{\omega_\ell} & = 
    \bra{\omega_i} \bra{\omega_k} (\mathcal{I} \otimes O)D[\mathcal{E}] \ket{\omega_j} \ket{\omega_\ell} +  \bra{\omega_i} \bra{\omega_k} D[\mathcal{E}] (\mathcal{I} \otimes O)\ket{\omega_j} \ket{\omega_\ell} \\ & =
    \bra{\omega_i} \lbrace O,\mathcal{E}(\ketbra{\omega_\ell}{\omega_k}) \rbrace \ket{\omega_j} \, ,
\end{split}
\end{equation}
where we have used the identity $\bra{\omega_k} D[\mathcal{E}] \ket{\omega_\ell} = \mathcal{E}(\ketbra{\omega_\ell}{\omega_k})$.
On the other hand. the r.h.s. reads
\begin{equation}
\begin{split}
     \bra{\omega_i} \bra{\omega_k} \lbrace \mathcal{I} \otimes \mathcal{E}^\dag (O),D[\mathcal{P}^\dag ] \rbrace \ket{\omega_j} \ket{\omega_\ell} & =   
      \bra{\omega_i} \bra{\omega_k} (\mathcal{I}\otimes \mathcal{E}^\dag (O))D[\mathcal{P}^\dag] \ket{\omega_j} \ket{\omega_\ell} +  \bra{\omega_i} \bra{\omega_k} D[\mathcal{P}^\dag] (\mathcal{I} \otimes \mathcal{E}^\dag (O))\ket{\omega_j} \ket{\omega_\ell} \\ & =
      (q_k + q_\ell)  \bra{\omega_i} \bra{\omega_k} D[\mathcal{P}^\dag ] \ket{\omega_j} \ket{\omega_\ell} 
      = (q_k + q_\ell)  \bra{\omega_i} \mathcal{P}^\dag (\ketbra{\omega_\ell}{\omega_k})  \ket{\omega_j} \, .
\end{split}
\end{equation}
Putting everything together and further assuming that $q_k + q_\ell \neq 0$ for all $k$ and $\ell$ we finally get to
\begin{equation}
  \mathcal{P}^\dag (\ketbra{\omega_k}{\omega_\ell}) = (q_k + q_\ell)^{-1} \lbrace O,\mathcal{E}(\ketbra{\omega_k}{\omega_\ell}) \rbrace \, .
\end{equation}

\section{Recovery map for unitary quantum channels}
\label{appendix_invertible_channel_recovery}
In this Appendix we explicitly show that, when the quantum channel $\mathcal{E}$ models a unitary evolution, namely $\mathcal{E}(\rho)=U\rho U^\dag$ with $UU^\dag = \mathcal{I}$, then its inverse $\mathcal{E}^{-1}(\rho)=U^\dag \rho U$
is a recovery map $\mathcal{P}$ for the pre-processing protocol, i.e. it solves
\begin{equation}
   \lbrace \mathcal{I} \otimes O , D[\mathcal{E}]\rbrace = \lbrace \mathcal{I} \otimes  \mathcal{E}^\dagger (O),D[{\mathcal{P}^\dag}]\rbrace \, .
    \label{appendix_bayes_rule_pre_processing}
\end{equation}
We showed in Section \ref{section_pre_processing} that the solution to Eq.~\eqref{appendix_bayes_rule_pre_processing} is given by
\begin{equation}
  \mathcal{P}^\dag (\ketbra{\omega_k}{\omega_\ell}) = (q_k + q_\ell)^{-1} \lbrace O,\mathcal{E}(\ketbra{\omega_k}{\omega_\ell}) \rbrace \, ,
  \label{app_eq_solu_pre_proc}
\end{equation}
where $q_k$ and $\ket{\omega_k}$ are the eigenvalues and eigenstates of $\mathcal{E}^\dag (O)$, respectively.
%It is well known that an invertible CPTP map $\mathcal{E}$ corresponds to a unitary evolution, i.e. $\mathcal{E}(\sigma) = U\sigma U^\dag$, with $U^\dag U = \mathcal{I}$.
The noisy reference observable reads
\begin{equation}
    \mathcal{E}^\dag (O) =  U^\dag O U = \sum_k q_k U^\dag \ketbra{\lambda_k} U \, ,
\end{equation}
where we have used the eigen-decomposition of the target observable, namely $O=\sum_k q_k \ketbra{\lambda_k}$.
It thus follows that $\ket{\omega_k} = U^\dag \ket{\lambda_k}$.
We can use all of these facts to rewrite Eq.~\eqref{app_eq_solu_pre_proc} as
\begin{equation}
\begin{split}
     \mathcal{P}^\dag (\ketbra{\omega_k}{\omega_\ell}) & = (q_k + q_\ell)^{-1} \lbrace O,\mathcal{E}(\ketbra{\omega_k}{\omega_\ell}) \rbrace \\ &  = (q_k + q_\ell)^{-1} \lbrace \textstyle\sum_i q_i \ketbra{\lambda_i} , U \ketbra{\omega_k}{\omega_\ell} U^\dag \rbrace 
     \\ & = (q_k + q_\ell)^{-1} \textstyle\sum_i q_i  \lbrace\ketbra{\lambda_i} , U U^\dag \ketbra{\lambda_k}{\lambda_\ell} U U^\dag \rbrace
     \\ & =  (q_k + q_\ell)^{-1} (q_k \ketbra{\lambda_k}{\lambda_\ell}+ q_\ell \ketbra{\lambda_k}{\lambda_\ell}) 
     \\ & = \ketbra{\lambda_k}{\lambda_\ell} = U \ketbra{\omega_k}{\omega_\ell} U^\dag \, , 
\end{split}
\end{equation}
which implies that $P^\dag = \mathcal{E}$, or equivalently that $\mathcal{P} = \mathcal{E}^\dag = \mathcal{E}^{-1}$, concluding our proof.
%Without $\tau$ in Eq.~\eqref{appendix_bayes_rule_pre_processing}, the expression for the recovery map given by Eq.~\eqref{app_eq_solu_pre_proc} would have $\mathcal{E}^\dag$ instead of $\mathcal{E}$ and the recovery map $\mathcal{R}$ would be different from the inverse noise map \GB{(This last sentence is not true. Without the $\tau$ we would get the same equation and same solution (QOOT is Hermitian)}.

\section{Post-processing protocol recovery map (qubit systems)}
\label{appendix_qubit_system}
Once again, the reader may use Figure \ref{fig_post_processing} as a reference for the spaces that maps and operators act upon.
In the main text we showed that a recovery map for a post-processing protocol may be obtained by solving 
\begin{equation}
        \lbrace O\otimes\mathcal{I},D[\mathcal{R}^\dag]\rbrace =  \lbrace  \mathcal{R}^\dag (O) \otimes \mathcal{I} , D[\mathcal{E}]\rbrace \, .
\end{equation}
The difficulty in solving the above equation lies in the fact that, unlike the pre-processing scenario, $\mathcal{R}^\dag$ appears on both sides of the equation and, more importantly, we do not know anything about $\mathcal{R}^\dag(O)$, let alone its diagonalization. Instead, we consider the eigen-decomposition of the reference observable $O = \sum_i q_i \ketbra{\omega_i}$ and  project both sides of the equation above on the joint Hilbert space $\mathcal{H}_A \otimes \mathcal{H}_B$ basis $\lbrace \ket{\omega_i}\otimes\ket{\omega_j} \rbrace \equiv \lbrace \ket{\omega_i}\ket{\omega_j} \rbrace$. 
In particular, the l.h.s. reads
\begin{equation}
\begin{split}
    \bra{\omega_k}\bra{\omega_i} \lbrace O\otimes\mathcal{I},D[\mathcal{R}^\dag]\rbrace \ket{\omega_\ell} \ket{\omega_j} & = \bra{\omega_k}\bra{\omega_i} ( O\otimes\mathcal{I})D[\mathcal{R}^\dag]\ket{\omega_\ell} \ket{\omega_j}  + \bra{\omega_k}\bra{\omega_i} D[\mathcal{R}^\dag]  (O\otimes\mathcal{I})\ket{\omega_\ell} \ket{\omega_j} \\
    & = (q_k + q_\ell)  \bra{\omega_k}\bra{\omega_i}  D[\mathcal{R}^\dag] \ket{\omega_\ell} \ket{\omega_j} = (q_k + q_\ell) \bra{\omega_i} \mathcal{R}^\dag(\ketbra{\omega_\ell}{\omega_k}) \ket{\omega_j} \, . 
\end{split}
\end{equation}
With similar techniques, we compute the r.h.s., namely
\begin{equation}
\begin{split}
    \bra{\omega_k}\bra{\omega_i} \lbrace \mathcal{R}^\dag(O) \otimes \mathcal{I}  , D[\mathcal{E}]\rbrace \ket{\omega_\ell} \ket{\omega_j} & = \bra{\omega_i} \lbrace \mathcal{R}^\dag(O),\mathcal{E}(\ketbra{\omega_\ell}{\omega_k}) \rbrace \ket{\omega_j} \, .
\end{split}
\end{equation}
Putting everything together, we obtain
\begin{equation}
    (q_k + q_\ell) \mathcal{R}^\dag(\ketbra{\omega_\ell}{\omega_k})  =  \lbrace \mathcal{R}^\dag(O),\mathcal{E}(\ketbra{\omega_\ell}{\omega_k}) \rbrace \, .
    \label{matrix_elements_equation_1}
\end{equation}
Notice that, while we do not know what $\mathcal{R}^\dag(O)$ is, we may conveniently express it as $\mathcal{R}^\dag(O)=\sum_i q_i \mathcal{R}^\dag(\ketbra{\omega_i})$, by linearity. Hence, we can rewrite Eq.~\eqref{matrix_elements_equation_1} as 
\begin{equation}
    (q_k + q_\ell) \mathcal{R}^\dag(\ketbra{\omega_\ell}{\omega_k})  =  \sum_i q_i\lbrace \mathcal{R}^\dag(\ketbra{\omega_i}),\mathcal{E}(\ketbra{\omega_\ell}{\omega_k}) \rbrace \, .
    \label{full_equation}
\end{equation}
The equation above can be recast as
\begin{align}
    X_{k\ell} = \sum_i \lbrace X_{ii},A_{k\ell} \rbrace \, ,
    \label{post_processing_recasted}
\end{align}
where 
\begin{equation}
    X_{k\ell} = (q_k + q_\ell)  \mathcal{R}^\dag (\ketbra{\omega_k}{\omega_\ell}) \, ,
\end{equation}
\begin{equation}
    A_{k\ell}  = \frac{1}{2}\mathcal{E}(\ketbra{\omega_k}{\omega_\ell}) \, .
\end{equation}
All is left now, is to solve Eq.~\eqref{post_processing_recasted}. 
For a qubit system, we may decompose all operators appearing in the equation in the Pauli basis, namely
\begin{align}
    X_{k\ell} = \sum_{j=0}^3 x^{(k\ell)}_j \sigma_j \, , \quad A_{k\ell} = \sum_{j=0}^3 a^{(k\ell)}_j \sigma_j \, ,
\end{align}
where $\lbrace \sigma_0,\sigma_1,\sigma_2,\sigma_3 \rbrace \equiv \lbrace\mathcal{I},X,Y,Z \rbrace$ denote the Pauli operators.
We then substitute the above expression into Eq.~\eqref{post_processing_recasted}, yielding
\begin{equation}
    \sum_{j} x^{(k\ell)}_j \sigma_j = \sum_{imn} x^{(ii)}_m a^{(k\ell)}_n \lbrace \sigma_m, \sigma_n \rbrace \, .
    \label{single_qubit_equation1}
\end{equation}
The anticommutator of two Pauli matrices equals to $\lbrace \sigma_m,\sigma_n \rbrace=2\mathcal{I}$ for every $m=n$, $\lbrace \sigma_0,\sigma_n \rbrace=2\sigma_n$, and is zero otherwise. The r.h.s. of Eq.~\eqref{single_qubit_equation1} then reads
\begin{equation}
    \sum_{imn} x^{(ii)}_m a^{(k\ell)}_n \lbrace \sigma_m, \sigma_n \rbrace = \sum_i \left[ \sum_n  x^{(ii)}_n a^{(k\ell)}_n 2\mathcal{I} + \sum_{n\neq 0} \left( x^{(ii)}_0 a^{(k\ell)}_n  +  x^{(ii)}_n a^{(k\ell)}_0  \right)2\sigma_n \right] \, .
\end{equation}
Putting everything together, we obtain
\begin{equation}
    \sum_j x^{(k\ell)}_j \sigma_j = \sum_i  \sum_n  x^{(ii)}_n a^{(k\ell)}_n 2\mathcal{I} + \sum_i \sum_{n\neq 0} \left( x^{(ii)}_0 a^{(k\ell)}_n  +  x^{(ii)}_n a^{(k\ell)}_0  \right)2\sigma_n \, , 
\end{equation}
which leads to the following set of linear equations for the (eight) coefficients 
$x_j^{(k\ell)}$
\begin{equation}
     x^{(k\ell)}_0  = 2 \sum_{i,j} x^{(ii)}_j a^{(k\ell)}_j \, ,
     %label{equivalent_equation1}
\end{equation}
\begin{equation}
    x^{(k\ell)}_j  = 2 \sum_{i} \left( x^{(ii)}_0 a^{(k\ell)}_j  +  x^{(ii)}_j a^{(k\ell)}_0  \right) \, , \quad j\in\lbrace 1,2,3 \rbrace \, .
    \label{equivalent_equation2}
\end{equation}

\section{Post-processing protocol recovery map ($d-$dimensional system)}
\label{appendix_d_dimensional_system}
We now want to generalize the expressions and the system of linear equations obtained in the previous section for a generic $d-$dimensional system.
Once again, the idea is to start from $X_{k\ell} = \sum_i \lbrace X_{ii}, A_{k\ell}\rbrace$ and decompose the operators appearing in the equation onto a suitable operator basis. 
To exploit the structure of the equation, we need a basis that has simple anti-commutation properties.
For $n-$qubit systems, we might be tempted to use the Pauli basis $\sigma_{j_1}\otimes \dots \otimes \sigma_{j_n}$ to decompose our operators, however it becomes tricky to concisely express the anticommutator of two basis elements. 
Hence, we resort to the basis constructed starting from the so called \textit{clock} and \textit{shift} operators.
In dimension $d=2$, these are simply the Pauli operators $\sigma_1$ and $\sigma_3$, and they satisfy $\sigma^2_1 = \sigma^2_3 = \mathcal{I}$ and $\sigma_1 \sigma_3 = -\sigma_3 \sigma_1 = e^{i\pi} \sigma_3 \sigma_ 1$. In arbitrary dimension $d$ ($d=2^n$ for $n$-qubit systems) these operators generalize to 
\begin{equation}
    \Sigma_1 = \begin{pmatrix}
        0 & 0 & 0 & \dots & 0 & 1 \\ 
        1 & 0 & 0 & \dots & 0 & 0 \\ 
        0 & 1 & 0 & \dots & 0 & 0 \\ 
        0 & 0 & 1 & \dots & 0 & 0 \\ 
        \vdots & \vdots & \vdots & \ddots & \vdots & \vdots \\ 
        0 & 0 & 0 & \dots & 1 & 0  
    \end{pmatrix} \, , \quad\quad 
    \Sigma_3 = \text{diag}(1,\omega,\omega^2,\dots,\omega^{d-1}) \, , 
\end{equation}
where $\omega=e^{\frac{2\pi i}{d}}$. 
These matrices are unitary, traceless but not Hermitian for $d>2$.
They satisfy $\Sigma_1^d = \Sigma_3^d=\mathcal{I}$ as well as $\Sigma_3 \Sigma_1 = \omega \Sigma_1 \Sigma_3$ and, most importantly, they may be used to define the following complete set of $d^2$ $d\times d$ matrices, namely
\begin{equation}
    \sigma_{j,k} = \Sigma_1^k \Sigma_3^j \, ,
\end{equation}
with $j,k \in \lbrace 1,\dots,d \rbrace$.
By repeatedly using the defining property $\Sigma_3 \Sigma_1 = \omega \Sigma_1 \Sigma_3$, one can also show that $\Sigma_3^j \Sigma_1^k = \omega^{jk}\Sigma_1^k \Sigma_3^j$.
It follows that
\begin{equation}
    \lbrace \sigma_{j,k},\sigma_{l,m}\rbrace = (\omega^{kl} + \omega^{mj})\sigma_{j+l,k+m} \, .
    \label{anti_commutation_relation_clock_shift}
\end{equation}
As a result, for a $d-$dimensional system we may decompose the operators $X_{k\ell}$ and $A_{k\ell}$ as follows 
\begin{equation}
    X_{k\ell}  = \sum_{l,m=1}^d x^{(k\ell)}_{lm} \sigma_{l,m} \, ,\quad
    A_{k\ell}  = \sum_{l,m=1}^d a^{(k\ell)}_{lm} \sigma_{l,m} \, .
\end{equation}
Substituting these expressions into $X_{k\ell} = \sum_i \lbrace X_{ii}, A_{k\ell}\rbrace$ and using the anti-commutation relation Eq.~\eqref{anti_commutation_relation_clock_shift} leads to 
\begin{equation}
    \sum_{lm} x^{(k\ell)}_{lm} \sigma_{l,m} = \sum_{istpn} x^{(ii)}_{st} a^{(k\ell)}_{pn} \lbrace \sigma_{s,t},\sigma_{p,n} \rbrace = \sum_{istpn} x^{(ii)}_{st} a^{(k\ell)}_{pn} (\omega^{tp}+\omega^{sn}) \sigma_{s+p,t+n} \, . 
\end{equation}
Consequently, we obtain the following linear equation for the coefficients $x^{(k\ell)}_{lm}$
\begin{equation}
    x^{(k\ell)}_{lm} = \sum_i \sum_{s+p=l} \sum_{t+n=m} x^{(ii)}_{st} a^{(k\ell)}_{pn} (\omega^{tp}+\omega^{sn}) \, .
\end{equation}

\section{}
\label{appendix_trace_contraint}
Here we prove that the  necessary and sufficient condition $\lbrace O,\mathcal{R}(\mathcal{I}) \rbrace = 2O$ for the Jordan product QOOT $\mathcal{R}^\dag \star O = \frac{1}{2}\lbrace O \otimes \mathcal{I},  D[\mathcal{R}^\dag]\rbrace$ to satisfy its defining property, is equivalent to imposing $\Tr[X_{k\ell}] = q_k \delta_{k\ell}$, where $X_{k\ell} = (q_k + q_\ell)  \mathcal{R}^\dag (\ketbra{\omega_k}{\omega_\ell})$.
We remind the reader that $q_k$ and $\ket{\omega_k}$ are the eigenvalues and the eigenstates of the reference observable $O$, respectively.
\begin{equation}
\begin{split}
    \frac{1}{2}\lbrace O,\mathcal{R}(\mathcal{I}) \rbrace & = \frac{1}{2} \Tr_B [\lbrace O\otimes\mathcal{I} , D[\mathcal{R}^\dag]\rbrace] = \frac{1}{2}\sum_{k\ell} \lbrace O,\ketbra{\omega_k}{\omega_\ell}\rbrace \Tr[\mathcal{R}^\dag (\ketbra{\omega_\ell}{\omega_k})]  =
    \frac{1}{2}\sum_{k\ell} (q_k+q_\ell) \ketbra{\omega_k}{\omega_\ell} \Tr[\mathcal{R}^\dag (\ketbra{\omega_\ell}{\omega_k})] \\ & = 
      \frac{1}{2}\sum_{k\ell}   \Tr[(q_k+q_\ell)\mathcal{R}^\dag (\ketbra{\omega_\ell}{\omega_k})]\ketbra{\omega_k}{\omega_\ell}  = 
     \frac{1}{2}\sum_{k\ell}   \Tr[X_{k\ell}]\ketbra{\omega_k}{\omega_\ell} \, , 
\end{split}
\end{equation}
where in the second equality we have exploited the definition of the Jamio\l{}kowski state Eq.~\eqref{jamiolkowsi_state} using $\lbrace \ket{\omega_k} \rbrace$ as the basis of choice.
The chain of identities above implies that imposing $\frac{1}{2}\lbrace O,\mathcal{R}(\mathcal{I}) \rbrace = O$ is equivalent to imposing $ \frac{1}{2}\sum_{k\ell}   \Tr[X_{k\ell}]\ketbra{\omega_k}{\omega_\ell} = \sum_k q_k \ketbra{\omega_k}$. Finally, this last constraint is clearly satisfied if an only if $\Tr[X_{k\ell}] = q_k \delta_{k\ell}$.

\end{document}